\documentclass[11pt,a4paper]{book}

\usepackage{amssymb}
\usepackage{latexsym}
\usepackage{multirow,booktabs}
\usepackage{threeparttable}
\usepackage{graphicx}
\usepackage{psfrag,graphicx}
\usepackage{pstcol,pst-text}
\usepackage{amsmath}
\usepackage{rotating,booktabs}

\usepackage{fancyhdr}

\def\mybibliography#1{{\begin{center} \bf References \end{center}}\list
 {}{\setlength{\leftmargin}{1em}\setlength{\labelsep}{0pt}
\itemindent=-\leftmargin}
 \def\newblock{\hskip .02em plus .20em minus -.07em}
 \sloppy\clubpenalty4000\widowpenalty4000
 \sfcode`\.=1000\relax}
\newbox\TempBox \newbox\TempBoxA
\def\uw#1{%
  \ifmmode\setbox\TempBox=\hbox{$#1$}\else\setbox\TempBox=\hbox{#1}\fi%
  \setbox\TempBoxA=\hbox to \wd\TempBox{\hss\char'176\hss}%
  \rlap{\copy\TempBox}\smash{\lower9pt\hbox{\copy\TempBoxA}}%
}
\newbox\TempBox \newbox\TempBoxA
\def\uwd#1{%
  \ifmmode\setbox\TempBox=\hbox{$#1$}\else\setbox\TempBox=\hbox{#1}\fi%
  \setbox\TempBoxA=\hbox to \wd\TempBox{\hss\char'176\hss}%
  \rlap{\copy\TempBox}\smash{\lower10pt\hbox{\copy\TempBoxA}}%
}
\def\mathunderaccent#1{\let\theaccent#1\mathpalette\putaccentunder}
\def\putaccentunder#1#2{\oalign{$#1#2$\crcr\hidewidth
\vbox to.2ex{\hbox{$#1\theaccent{}$}\vss}\hidewidth}}

\newcommand{\bee}{\begin{eqnarray*}}
\newcommand{\eee}{\end{eqnarray*}}
\newcommand{\be}{\begin{eqnarray}}
\newcommand{\ee}{\end{eqnarray}  }

\def\ni{\noindent}

\def\be{\begin{eqnarray*}}
\def\ee{\end{eqnarray*}}

\def \baselinestretch{1.5}

\begin{document}
\thispagestyle{empty}

\vspace{0.4pc}

\begin{center}
{\large \bf COX REGRESSION ANALYSYS WITH MISSING COVARIATES VIA NONPARAMETRIC MULTIPLE IMPUTATION}
\end{center}

\vspace{.2cm}
\begin{center}
\renewcommand{\thefootnote}{\fnsymbol{footnote}}
Chiu-Hsieh\ Hsu$^{1,2}$, Mandi\ Yu$^{3}$\\

{\it } $^{1}$ Department of Epidemiology and Biostatistics, Mel and Enid Zuckerman College of Public Health,
and $^{2}$ University of Arizona Cancer Center, Tucson, AZ, 85724, USA\\
$^{3}$ Surveillance Research Program, Division of Cancer Control and Population Sciences, National Cancer Institute, National Institutes of Health. Rockville, MD.\\
\end{center}

{\small
\begin{quotation}
\noindent {\it Abstract}: We consider the situation of estimating Cox regression in which some covariates are subject to missing, and there exists additional information (including observed event time, censoring indicator and fully observed covariates) which may be predictive of the missing covariates. We propose to use two working regression models: one for predicting the missing covariates and the other for predicting the missing probabilities. For each missing covariate observation, these two working models are used to define a nearest neighbor imputing set. This set is then used to nonparametrically impute covariate values for the missing observation. Upon the completion of imputation, Cox regression is performed on the multiply imputed datasets to estimate the regression coefficients. In a simulation study, we compare the nonparametric multiple imputation approach with the augmented inverse probability weighted (AIPW) method, which directly incorporates the two working models into estimation of Cox regression. We show that all approaches can reduce bias due to non-ignorable missing mechanism. The proposed nonparametric imputation method is robust to mis-specification of either one of the two working models and robust to mis-specification of the link function of the two working models. In contrast, the AIPW method is not robust to mis-specification of the link function of the two working models and is sensitive to the selection probability. We apply the approaches to a breast cancer dataset from Surveillance, Epidemiology and End Results (SEER) Program .

\noindent{\it keywords}: Augmented Inverse Probability Weighted Method; Cox Regression; Missing Covariates; Multiple Imputation\\
\end{quotation}
}

\noindent{\bf 1. Introduction}\\
For survival time data with covariates, Cox regression is often used to specify the relationship between survival time and covariates (Cox, 1972). For time-independent covariates, Cox regression has the proportional hazards property. It estimates the regression coefficients of the model using the partial likelihood function without specifying the baseline hazard function (Cox, 1975). The estimators of regression coefficients have been shown to be consistent, normally distributed and semiparametrically efficient (Andersen, 1982). However, in many situations, some of the covariates are not fully observed. Missing covariates could compromise the asymptotic properties of the estimators if missing data are not accounted for in estimation. Specifically, it has been shown that the estimators of the regression coefficients derived from the subjects with all of the covariates observed (i.e. complete-case analysis) not only lose efficiency, but may also generate biased regression coefficient estimates when missingness depends on the survival outcome (i.e. survival time and censoring indicator) (Little and Rubin, 2002).  When missingness depends on the survival outcome (i.e. survival time and censoring indicator) and some fully observed covariates, missing mechanism is considered as missing at random (MAR) (Rubin, 1987). For the survival outcome data, MAR can be even further classified into two scenarios: failure-ignorable MAR (i.e. missingness does not depend on failure time) and censoring-ignorable MAR (i.e. missingness does not depend on censoring time but may depend on failure time) (Rathouz, 2007). When missingness is failure-ignorable MAR, complete-case analysis can still produce valid regression coefficient estimates. However, when missingness is censoring-ignorable MAR, complete-case analysis may produce biased regression coefficient estimates.      

Several approaches have been proposed to deal with missing covariates in Cox regression. Of the existing approaches, the augmented inverse probability weighted (AIPW) method (Robins et al., 1994; Wang and Chen, 2001), where the weight is derived from a fully specified model for the missing status conditional on the observed data and an augmentation term derived from a fully specified model for the missing covariate conditional on the observed data is added to estimation to correct the potential bias, is a very popular method and has been shown to have a double robustness property. The AIPW method uses two fully specified parametric models to account for missing covariates while estimating the regression coefficients of Cox regression model.  This indicates that at least one of the two models has to be correctly specified, including the distribution and link function for the missing covariate and the missing status, respectively. To weaken the reliance on parametric assumptions behind the two models, non-parametric regression has been used to estimate the two models without fully specifying the relationship between the missing covariates and the observed data (Qi et al., 2005). As the dimensionality of the observed data increases, it becomes extremely difficult to use non-parametric regression to estimate the two models.   

We previously developed a nonparametric multiple imputation approach to deal with missing data in a situation without censored data (Long et al., 2012). The approach indirectly uses two working models to recover information for missing data observations. Specifically, we use two working regression models, one for predicting the missing covariate values and one for predicting the missing probabilities. The parameter estimates from these two working models are then used to give two predictive scores for each subject, defined as the linear combination of the covariates in the corresponding model. The method then selects an imputing set of observations for each missing data observation, which consists of subjects who have their data fully observed and have similar predictive scores as the subject with missing data. Then the missing data value is randomly drawn from this imputing set. The idea is similar to predictive mean matching (Rubin, 1986) and propensity score matching (Rosenbaum and Rubin, 1985) in the missing data literature. In a situation with missing outcome data, we have shown this nonparametric multiple imputation approach can generate a consistent mean estimator.  To weaken the reliance on the two models, in this paper we generalize the nonparametric multiple imputation approach to handle estimation of Cox regression with missing covariates. Specifically, we propose to use two working regression models, one for predicting the missing covariates and one for predicting the missing probabilities, to derive two predictive scores to select an imputing set for each missing covariate observation. It has been shown that the survival outcome data (specifically cumulative baseline hazard and censoring indicator) need to be included in predicting the missing covariates (White and Royston, 2009). In addition, the survival outcome data can be also included in the regression model for missing probabilities as the covariates to account for potentially censoring-ignorable MAR. The two working regression models are only used to derive two predictive scores to select an imputing set. Hence, the approach can easily handle the multi-dimensional structure of the observed data and is expected to be less affected by the mis-specification of the two working models. In this paper, not only will we study the performance of the proposed multiple imputation approach but will also compare its performance with the AIPW method.

This paper is organized as follows. In Section 2, we review the complete-case analysis and the AIPW method. In Section 3, we describe the proposed multiple imputation method and the associated properties. In Section 4, we apply the techniques to data from a breast cancer study. In Section 5, we give results from a simulation study. A discussion follows in Section 6.

\noindent{\bf 2. Review of Methods}\\
In this section, we begin with describing the setting of the situation: estimation of Cox regression with time-independent covariates and one of the covariates subject to missing. Let $T$ denote the failure time, $C$ denote the censoring time, $Y=min(T,C)$ denote the observed time, $\delta_t=I[T\le C]$ denote the censoring indicator and $N(t)=\delta_t I(T \le t)$ denote the counting process. Assume $T$ has a hazard function of $\lambda(t)=\lambda_0(t)exp(\beta_x *X+\beta_z*Z)$, where $\lambda_0(t)$ is an unspecified baseline hazard function, $X$ is subject to missing and $Z$ is fully observed. Let $\delta_x$ denote the missing indicator for $X$ (i.e. $\delta_x=1$ if $X$ is observed; otherwise, 0) and $\pi=Pr(\delta_x=1)$ denote the selection probability. We assume that $T$ and $C$ are independent conditional on $X$, $Z$ and $X$ is missing at random (i.e. $E[\delta_x|Y,\delta_t,Z,X]=E[\delta_x|Y,\delta_t,Z]$) and there is a random sample of $n$ subjects.

\ni{\it Complete-case analysis}\\
The complete-case (CC) analysis of $\beta=(\beta_x,\beta_z)$ is based on the partial likelihood estimator using observations that have $X$ observed. Let $r_i(\beta, t)=exp[\beta_x X_i+\beta_z Z_i] \equiv r_i^{(0)}(\beta,t)$ and $r_i^{(1)}=(X_i Z_i)^{'}r_i(\beta,t)$. The CC analysis involves solving the following estimating equations:
$$U_{cc}(\beta)=\sum_{i=1}^{n} \left [\delta_{t_i}\delta_{x_i}\left\{ 
\left( \begin{array}{c} X_i\\ Z_i\\ \end{array} \right)
- \frac{S_{cc}^{(1)}(\beta, T_i)}{S_{cc}^{(0)} (\beta, T_i)}\right\}\right]=0,$$ 
where $S_{cc}^{(m)}(\beta, T_i)=n^{-1}\sum_{j=1}^{n} \delta_{x_i} I(T_j \ge T_i)r_j^{(m)}(\beta, T_i)$ for $m=0,1$. It is easy to implement the CC analysis and it is consistent when the missingness depends only on $Z$. However, it loses efficiency due to discarding data from incomplete observations, especially when the missing rate is greater than 25\% (Marshall et al., 2010), and is inconsistent when missingness depends on $T$ or $\delta_t$. 

\ni{\it AIPW method}\\
The AIPW method was first proposed by (Robins et al., 1994) to modify the CC analysis to produce consistent estimators of $\beta$ and furthermore improve efficiency of the CC analysis. The AIPW method has been studied and further developed by a few groups for various scenarios. For Cox regression with a missing covariate, it involves solving the estimating equations (Wang and Chen, 2001; Qi et al., 2005):
$$U_{AIPW}(\beta)=\sum_{i=1}^{n} \left [\frac{\delta_{t_i}\delta_{x_i}}{\pi_i}\left \{ 
\left( \begin{array}{c} X_i\\ Z_i\\ \end{array} \right)
- \frac{S_{AIPW}^{(1)}(\beta, T_i)}{S_{AIPW}^{(0)} (\beta, T_i)} \right \}+A_i(\beta,\pi) \right ]=0,$$ 
where 
\begin{equation*}
    \begin{split}
S_{AIPW}^{(m)}(\beta, T_i) &= \\
&n^{-1}\sum_{j=1}^{n} \{\frac{ \delta_{x_j}}{\pi_j} I(T_j \ge T_i)r_j^{(m)}(\beta, T_i) +(1-\frac{\delta_{x_j}} {\pi_j}) I(T_j \ge T_i)E[r_j^{(m)}(\beta,T_i)|T_i,\delta_{t_i}, Z_i]\}
    \end{split}
\end{equation*}
for $m=0,1$
 and 
 \begin{equation*}
 \begin{split}
 A_i(\beta,\pi) &=\\
 &\left (1-\frac{\delta_{x_i}}{\pi_i} \right ) \int_0^\tau \left \{E \left [ \left( \begin{array}{c} X_i\\ Z_i\\ \end{array}\right) [dN_i(t)|Y_i,\delta_i,Z_i]-\frac{S_{AIPW}^{(1)}(\beta, T_i)}{S_{AIPW}^{(0)} (\beta, T_i)} E[dN_i(t)|Y_i,\delta_{t_i},Z_i \right ]\right \}.
 \end{split}
 \end{equation*}
Based on the above expression, it can be seen that the conditional expectation in $A_i(\beta,\pi)$ depends on the baseline cumulative hazard and the conditional distribution of $X|T,\delta_t,Z$. The EM algorithm can be used to derive the AIPW estimates (Wang and Chen, 2001). To perform the EM algorithm, the conditional distribution of $X|T,\delta_t, Z$ and the selection probability $\pi$ need to be estimated. It has been shown that if one of them is estimated correctly, the AIPW estimator is consistent (so called double robustness property). Often two parametric working models are used to estimate the conditional distribution of $X|T,\delta_t,Z$ and the selection probability $\pi$, respectively, and then directly incorporate them into estimation of AIPW estimator.  To relax the reliance on the distributional assumptions, nonparametric techniques have been proposed to estimate the conditional distribution and the selection probability. However, as the number of fully observed covariates (i.e. $Z$) increases, it gets difficult to estimate the conditional distribution and the selection probability nonparametrically. In this paper, we will mainly focus on the performance of the AIPW estimator where two parametric working models are used to estimate the conditional distribution and the selection probability, respectively, and one of the two models is mis-specified. The estimate of standard error for AIPW is derived from 500 bootstrap samples.  

\noindent{\bf 3. Nonparametric Multiple Imputation}\\
Instead of directly incorporating the working models into estimation, we propose to use two working regression models, one for predicting the missing covariates and one for predicting the missing probabilities, to derive two predictive scores to select an imputing set for each missing covariate observation. The two working regression models are only used to derive two predictive scores to select an imputing set. Hence, the approach is expected to be less affected by the mis-specification of the two working models. To conduct nonparametric multiple imputation, for each missing covariate observation we seek an imputing set consisting of subjects who have similar predictive scores as the subject with missing covariate observation. We describe the imputation procedures in detail below.

\noindent{\it Imputation procedures for missing covariate $X$}\\
\noindent {\it Step 1: Estimate the two predictive scores on a Bootstrap sample}

\noindent To define each imputing set, we first reduce the observed survival data and $Z$ to two scalar indices (predictive scores), which provide an indicator of an individual's value of $X$ and chance of having missing $X$.  It has been shown that the conditional distribution of $X|T,\delta_t,Z$ depends on cumulative baseline hazard $H_0(t)$, censoring indicator $\delta_t$ and the fully observed covariate $Z$ (White and Royston, 2009). Hence, all of them will be included in the working regression model for predicting $X$. To account for potential censoring-ignorable MAR, we will include the survival outcome data (i.e. $Y$ and $\delta_t$), as well as $Z$, in the working regression model for predicting the missing probabilities. This strategy summarizes the multi-dimensional structure of the observed survival data and $Z$ into a two-dimensional summary. The hope is that this 2-dimensional summary contains most, if not all, the information about the value of missing $X$ and missingness. 

Specifically, a linear/generalized linear model with $H_0(t)$, $\delta_t$ and $Z$ as the covariates can be fitted to the complete cases to derive a predictive score for $X$. This score summarizes the relationship between $X$ and $H_0(t)$, $\delta_t$ and $Z$. A logistic regression model with the observed $Y$, $\delta_t$ and $Z$ as the covariates will be fitted to the missing indicator data (i.e. $\delta_x$) to derive a predictive score for missingness. This score summarizes the relationship between missingness and $Y$, $\delta_t$ and $Z$. The two models will be fitted on a nonparametric bootstrap sample (Efron, 1979) of the original dataset to incorporate the uncertainty of parameter estimates from the working models. This step results in proper multiple imputation (Nielsen, 2003 and references therein). More specifically, let $(Y^B, \delta_t^B, \delta_x^B, Z^B)$ denote the bootstrap sample. Two working models are conducted on the bootstrap sample to calculate two predictive scores, $S_x^{(B)}$ and $S_{\delta_x}^{(B)}$, for each individual in the bootstrap sample. We further standardize these scores by subtracting their sample mean and dividing by their standard deviation, and denote the standardized scores by $S_x^{c(B)}$  and $S_{\delta_x}^{c(B)}$, respectively. Combinations of these two predictive scores will be studied to see to what extent a double robustness property (Robins et al., 2000) for model mis-specification can be established and whether a robustness property for link function mis-specification can be established for the non-parametric multiple imputation method. 

\noindent {\it Step 2: Define the imputing set}

\noindent For subject $j$ with missing $X$ in the original dataset, two predictive scores are derived using the regression coefficient estimates obtained from the bootstrap sample (i.e. $S_x(j)$ and $S_{\delta_x}(j)$) and then standardized by subtracting the sample mean of the corresponding bootstrap sample predictive scores and dividing by the standard deviation of the corresponding bootstrap sample predictive scores, respectively (denoted as $S_x^c(j)$ and $S_{\delta_x}^c(j)$). The distance between subject $j$ in the original dataset and subject $k$ in the bootstrap sample is then defined as $d(j,k)=\sqrt{w_1 [S_x^c(j)-S_x^c(k) ]^2+w_2 [S_{\delta_x}^c(j)-S_{\delta_t}^c(k)]^2}$, where $w_1$ and $w_2$ are non-negative weights that sum to one. Non-zero weights for $w_2$ may be useful in reducing the bias resulting from model mis-specification. Specifically, a small weight $w_2$ (e.g. $0.2$) will result in incorporating the predictive scores from the missing probability model into defining a set of nearest neighbors for subjects with missing $X$. For subject $j$, the distance is then employed to define a set of nearest neighbors. This neighborhood consists of $NN$ subjects who have their $X$ observed and have a small distance from subject $j$ in terms of two predictive scores. 

\noindent {\it Step 3: Impute a value from the imputing set}

\noindent After the imputing set is defined, a value of $X$ is randomly drawn from the imputing set. Thus, the procedure imputes $X$ only from the subjects with $X$ observed. The non-parametric multiple imputation method based on a nearest neighborhood is denoted as $NNMI(NN,w_1,w_2)$. 

\noindent {\it Step 4: Repeat Steps 1 to 3 independently M times}

\noindent  Each of the $M$ imputed datasets is based on a different Bootstrap sample. Once the $M$ multiply imputed datasets are obtained, we carry out the MI analysis procedure established in (Rubin, 1987). Specifically for our purposes, Cox regression analysis with $X$ and $Z$ as the covariates is performed on the $M$ imputed datasets to estimate $\beta_x$ and $\beta_z$. The final estimate of $\beta_x$/$\beta_z$ is the average of the $M$ corresponding regression coefficient estimates and the final variance (denoted as $var[\hat \beta_x]$/ $var[\hat \beta_z]$) is the sum of the sample variance (denoted as $B_{\beta_x}$/$B_{\beta_z}$) of the $M$ regression coefficient estimates and the average (denoted as $U_{\beta_x}$/$U_{\beta_z}$) of the $M$ variance estimates of $\hat \beta_x$/$\hat \beta_x$. For both $\beta_x$ and $\beta_z$, the quantity $[\hat \beta-\beta]/\sqrt{var[\hat \beta]}$ approximately follows a $t$ distribution with a degree of freedom $v = (M-1)*[1 + \{U_{\beta}*M/(M+ 1)\}/B_{\beta}]^2$ (Rubin, 1987). We use a value of $10$ or higher for M.

\noindent{\bf 4. Illustration of the method on a breast cancer dataset }\\
We demonstrate the nonparametric multiple imputation approach on a dataset which consists of 7050 women diagnosed with stage IV breast cancer between 2005 and 2011 in California. This dataset was extracted from the breast cancer registries under Surveillance, Epidemiology and End Results (SEER) Program. Of the 7050, besides survival data (i.e. survival status and survival time) after diagnosed with breast cancer, for each patient there are several variables collected at diagnosis, as well as Age, Race (Black, White, Other), HER2, Radiation and Surgery. Those variables are summarized in Table 1. According to Table 1, of the 7050, 1293 (18.34\%) had missing HER2 value. Table 2 identifies variables predictive of HER2 value and missing probability. Specifically, based on univariate logistic regression analysis for HER2 positive indicator using patients with their HER2 value available (i.e. complete case analysis), Age, Race, Surgery and baseline cumulative hazard, respectively, are predictive of HER2 value. Based on univariate logistic regression for missing indicator, Age, Surgery, Radiation, survival status (Dead indicator) and baseline cumulative hazard, respectively, are predictive of missing probability. Those predictive covariates are then used to derive the conditional distribution of HER2 given the observed data and the selection probability for performing the AIPW estimation and derive two predictive scores for conducting the proposed multiple imputation method. Specifically, a working logistic regression model for HER2 positive indicator with Age, Race and Surgery, as well as survival status and baseline cumulative hazard, as covariates is fitted to derive the conditional distribution of HER2 given the observed data and a HER2 predictive score for each patient. A working logistic regression model for HER2 missing indicator with Age, Radiation and Surgery, as well as survival status and baseline cumulative hazard, as covariates is fitted to derive the selection probability (i.e. $\pi$=1-missing probability) and a predictive score of HER2 missing probability for each patient. To perform the AIPW estimation, the derived conditional distribution of HER2 is then used to derive the conditional expectations and the selection probability is incorporated into the estimation as the weight. To conduct the proposed multiple imputation approach (i.e. NNMI), the two predictive scores are then used to calculate the distance between patients and then select an imputing set for each patient with missing HER2. The number of imputes M is set at 50. Upon the completion of multiple imputation, Cox regression analysis with Age, Black and Others (White as the reference group), HER2, Radiation and Surgery as the covariates is performed on each of the imputed datasets and Rubin's rule (Rubin, 1987) is applied to derive the final estimate for each regression coefficient.

The results of the Cox regression estimation for the CC, AIPW and NNMI methods are provided in Table 3. Table 3 displays the hazard ratio estimate of each covariate along with the associated 95\% confidence interval (CI) and p-value. The CC and AIPW methods produce similar results. The results indicate that Age, Black and Surgery are significantly associated with survival after diagnosis with stage IV breast cancer. Specifically, older patients tend to have a higher hazard rate than younger patients, Black patients tend to have a higher hazard rate than White patients and patients without surgery tend to have a higher hazard rate than patients with surgery. Others patients have a slightly lower hazard rate than white patients but not significant at a significance level of 5\%. Radiation and HER2 are not significantly associated with survival after diagnosis with stage IV breast cancer. The NNMI method produces similar results as the CC and AIPW methods, except for Others. The result of NNMI method indicates that Others patients have a significantly lower hazard rate than White patients. In addition, the NNMI method produces a tighter 95\% CI than the CC and AIPW method except for HER2.     

\noindent{\bf 5. Simulation Study}\\
We perform several simulation studies to investigate the properties of the AIPW and NNMI methods when Cox regression has a covariate subject to missing and an additional fully observed covariate that is predictive of the missing covariate, and the quantities of interest are the regression coefficients of the Cox regression model. We investigate the effects of sample size, mis-specification of one of the two working models and mis-specification of the two link functions under a situation with dependent censoring. The simulation program is written in R and is available upon request. 

For each of 500 independent simulated datasets, the predictive covariate $Z$ is generated from a $U(0,1)$ distribution. The binary covariate $X$ subject to missing is generated from a $Bernoulli[p(Z)]$ distribution, where $p(Z)$ is either a constant or based on a logit link (i.e. $p(Z)=\frac {1}{1+e^{\alpha_0+\alpha_1 Z}}$ or a complementary log-log link (i.e. $p(Z)=e^{-e^{\alpha_0+\alpha_1 Z}}$). The failure time $T$ is generated from an exponential distribution with a hazard rate of $e^{\beta_x*X+\beta_z*Z}$. The censoring time C is generated from an exponential distribution with a hazard rate of $e^{\theta_x*X+\theta_z*Z}$. Let $Y=min(T,C)$ and $\delta_t=I(T \le C)$. The missing indicator $\delta_x$ ($\delta_x=1$ if $X$ is observed) is generated from a $Bernoulli[p(Z,Y)]$ distribution, where $p(Z,Y)$ (i.e. selection probability) is based on a logit link (i.e. $p(Z,Y)=\frac{1}{1+e^{\eta_0+\eta_zZ+\eta_y Y}}$) or a complementary log-log link (i.e. $p(Z,Y)=e^{-e^{\eta_0+\eta_zZ+\eta_y Y}}$). The regression coefficients and hazard rates are selected to give a desired censoring rate and missing rate.  

For the ``Fully-Observed" (FO) analysis, treated as the gold standard, we derive Cox regression coefficient estimates for each simulated dataset before any missingness is applied. For the ``Complete-Case" (CC) analysis, we derive Cox regression coefficient estimates from the data with $X$ observed. For the AIPW and NNMI methods, a working logistic regression model (denoted by $M_1$) is fitted to the data with $X$ observed to derive the conditional distribution of $X$ given the observed data and the predictive score of $X$. A working logistic regression model (denoted by $M_2$) is fitted to the missing indicator to derive the missing probability and the predictive score of missingness. When both working models include all of the correct covariates in the models (i.e. $M_1$: $Z$, $\delta_t$, $\hat H_0(t)$; $M_2$: $Z$, $Y$), they are denoted by AIPW$_{11}$ and NNMI$_{11}$, respectively. When the working model for predicting $X$ includes all of the correct covariates but the working model for predicting the missing probability does not (i.e. $M_1$: $Z$, $\delta_t$, $\hat H_0(t)$; $M_2$: $Z$), they are denoted by AIPW$_{12}$ and NNMI$_{12}$, respectively. When the working model for predicting $X$ does not include all of the correct covariates but the working model for predicting the missing probability does (i.e. $M_1$: $Z$, $\delta_t$; $M_2$: $Z$, $Y$), they are denoted by AIPW$_{21}$ and NNMI$_{21}$, respectively. When $X$ and $\delta_x$ are generated from a complementary log-log model, both AIPW and NNMI methods are considered as mis-specified even if both working models include all of the correct covariates in the models (i.e. AIPW$_{11}$ and NNMI$_{11}$) since the true models are not logit models. Based on our prior experience on dealing with missing data for survival analysis using multiple imputation, for the NNMI method we set $M=10$, $NN=5$, $w_1=0.8$ and $w_2=0.2$
 
The results are provided in Tables 4 and 5. The FO analysis, which is the gold standard method, targets the true values in all situations and produces coverage rates comparable to the nominal level, 95\%. The CC analysis as expected produces biased regression coefficient estimates in all situations and has a slightly lower coverage rate than the nominal level in some situations due to the bias. 

In all situations, both AIPW and NNMI methods produce reasonable regression coefficient estimates and coverage rates, for AIPW$_{11}$ and NNMI$_{11}$, i.e. when both working models include all of the correct covariates, and adequate performance if some covariates are omitted. For both $X$ independent (Table 4) and dependent (Table 5) of $Z$, when the working logistic regression model for $X$ (i.e. $M_1$) is misspecified (i.e. AIPW$_{21}$ and NNMI$_{21}$), the NNMI method has a larger bias, especially when the sample size is equal to 200. The bias decreases with sample size in all situations.  

The performance of the AIPW method highly depends on whether a correct model is used to derive the selection probability. In all situations when a correct model is used to derive the selection probability (i.e. AIPW$_{11}$ and AIPW$_{21}$, AIPW has a smaller bias and the coverage rate comparable to the nominal level. When a wrong model is used to derive the selection probability (i.e. AIPW$_{12}$), AIPW has a larger bias and a higher divergence rate in all situations.  Also, in all situations, when a wrong model is used to derive the selection probability, AIPW's standard errors tends to overestimate the variability of the regression coefficient estimates, and the overestimate is substantial. As a result, AIPW's coverage rates are higher than the nominal level even when the bias is larger. When the sample size is small, the AIPW method has a bias slightly smaller than the NNMI method except when the selection probability model is mis-specified. However, when the sample size increases to 400, both AIPW and NNMI methods have a similar bias. When $X$ is dependent of $Z$ (Table 5), the NNMI method is more efficient than the AIPW method as seen by the smaller SD and MSE values. 

In summary, all methods reduced the bias of the standard CC analysis, but the amount of the remaining bias, the efficiency and the validity of the estimated standard errors varied between methods. The performance of the AIPW method depends on whether a correct model is used to derive the selection probability. In contrast, the NNMI method in which two predictive scores are derived from two working regression models can provide reasonable regression coefficient estimates for both $X$ independent and dependent of $Z$  and is robust to mis-specification of either one of the two working regression models.

\noindent{\bf 6. Discussion}\\
In this paper we propose a nonparametric multiple imputation approach to handle a missing covariate in Cox regression analysis and compare it with an existing popular AIPW approach. Based on the simulation results, the performance of the AIPW method, depends on whether the selection/missing probability model is correctly specified. This indicates while performing the AIPW method, one has to be sure the corresponding model is correct, and specifically requires all aspects of the models including the link functions and choice of covariates to be correct.  In contrast, for the nonparametric multiple imputation approach the two working regression models are only used to derive two predictive scores to select imputing sets for missing covariate observations. Once the imputing sets are selected, nonparametric multiple imputation procedures are conducted on the sets. Therefore, this approach is expected to have weak reliance on the two working regression models compared to the AIPW method. 

The performances of the proposed nonparametric multiple imputation method will depend on the missing rate. Specifically, the missing rate will affect the number of similar ``donors" for each missing covariate observation. In a situation with a high missing rate, say, 0.90, a much larger sample size is required for the proposed method to perform well, than a situation with a low missing rate.
   
In this paper, we assume missingness only depends on the observed data.  This assumption is untestable. It is possible that missingness also depends on some unobserved data. This indicates non-ignorable missing mechanism may still remain even conditioning on all of the observed data.  Sensitivity analysis (Carpenter et al., 2007) would be a possible way to evaluate the impact of  unobserved data on the proposed multiple imputation approaches.

\noindent{\bf   References}
\begin{description}
\item Andersen, P. K. and Gill, R. D. (1982) Cox’s regression model for counting processes: a large sample study. \textit{Annals of Statistics}; 10: 1100–1120.

\item Carpenter, J. R., Kenward, M. G. and White, I. R. (2007) Sensitivity analysis after multiple imputation under missing at random: a weighting approach. {\it Statistical Methods in Medical Research}; \textbf{16}: 259-275.

\item Cox, D. R. (1972) Regression models and life-tables. \textit{Journal of the Royal Statistical Society. Series B (Methodological)}; \textbf{34}: 187-220. 

\item Cox, D. R. (1975) Partial likelihood. \textit{Biometrika}; 62: 269–276.

\item Efron, B. (1979) Bootstrap methods: another look at the jackknife. {\it Annals of Statistics}; {\bf 7}:1-26.

\item Little, R. J. A. and Rubin, D. B. (2002) {\it Statistical analysis with missing data} (2nd edn). Wiley: New York.

\item Long, Q., Hsu, C.-H. and Li, Y. (2012) Doubly robust nonparametric multiple impuatation for ignorable missing data.
{\it Statistica Sinica}; {\bf 22}: 149-172.

\item Marshall, A., Altman, D. G., Royston, P. and Holder, R. L. (2010) Comparison of techniques for handling missing covariate data within prognostic modelling studies: a simulation study. {\it BMC Medical Research Methodology}; \textbf{10}: 7. 

\item Nielsen, S. F. (2003) Proper and improper multiple imputation. {\it International Statistical Review}; {\bf 71}: 593607.

\item Qi, L., Wang, C. Y. and Prentice, R. L. (2005) Weighted estimators for proportional hazards regression with missing covariates. {\it Journal of the American Statistical Association}; {\bf 100}: 1250-1263.

\item Rathouz, P. J. (2007) Identifiability assumptions for missing covariate data in failure time regression models.
\textit{Biostatistics}; \textbf{8}: 345-356.

\item Robins, J. M., Rotnitzky, A. and Zhao, L. P. (1994) Estimation of regression coefficients when some regressors are not always observed. \textit{Journal of the American Statistical Association}; \textbf{89}: 846-866.

\item Robins, J. M., Rotnitzky, A. and van der Laan, M. (2000) Comment on On profile likelihood. {\it Journal of the American Statistical Association}; {\bf 95}:477 482.

\item Rosenbaum, P. R. and Rubin, D. B. (1985) Constructing a control group using multivariate matched sampling methods that incorporate the propensity score. \textit{American Statistician}; \textbf{39}, 33--38.

\item Rubin, D. B. (1986) Statistical matching using file concatenation with adjusted weights and multiple imputation.
\textit{Journal of Business \& Economic Statistics}; \textbf{4}, 87--94.

\item Rubin, D. B.   (1987) \textit{Multiple Imputation for Nonresponse in Surveys}. Wiley: New York.

\item Wang, C. Y. and Chen, H. Y. (2001) Augmented inverse probability weighted estimator for Cox missing covariate regression. 
{\it Biometrics}; {\bf 57}: 414-419.

\item White, I. R. and Royston, P. (2009) Imputing missing covariate values for the Cox model
{\it Statistics in Medicine}; \textbf{28}: 1982-1998.
\end{description}
\def\baselinestretch{1.0}  
\newpage
\begin{table}
\caption{Data Analysis: Description of the 7050 stage IV breast cancer patients}
\begin{center}
\begin{tabular}{lcc}
\hline
Variable & mean/frequency & Standard deviation/percentage\\
\hline
Age & 60.91 & 14.41\\
Race & &\\
\multicolumn{1}{r} {White} & 5585 & 79.22\\
\multicolumn{1}{r} {Black} & 721 & 10.23\\
\multicolumn{1}{r} {Others} & 744 & 10.55\\
HER2 & &\\
\multicolumn{1}{r} {Negative} & 4180 & 59.29\\
\multicolumn{1}{r} {Positive} & 1577 & 22.37\\
\multicolumn{1}{r} {Missing} & 1293 & 18.34\\
Surgery & &\\
\multicolumn{1}{r} {No} & 3916 & 55.55\\
\multicolumn{1}{r} {Yes} & 3134 & 44.45\\
Radiation & &\\
\multicolumn{1}{r} {No} & 4484 & 63.60\\
\multicolumn{1}{r} {Yes} & 2566 & 36.40\\
\hline
\end{tabular}
\end{center}
\end{table}

\begin{table}
\caption{Data Analysis: Identification of factors associated with missing value and probability of HER2}
\begin{center}
\begin{tabular}{lcccccc}
\hline
     & \multicolumn{3}{c}{Missing Value} & \multicolumn{3}{c}{Missing Probability}\\
\hline
Variable & OR$^a$ & 95\% CI$^b$ & p$^c$  & OR & 95\% CI & p \\
\hline
Age & 0.987 & (0.983, 0.991) & $<$0.0001 & 1.031 & (1.026, 1.035) & $<$0.0001\\
Black & 1.187 & (0.983, 1.433) & 0.08 & 1.007 & (0.825,1.229) & 0.94\\
Other & 1.360 & (1.135, 1.629) & $<$0.001 & 0.908 & (0.742, 1.112) & 0.35\\
No Radiation & 0.913 & (0.811,1.028) & 0.13 & 1.580 & (1.385, 1.804) & $<$0.0001\\
No Surgery & 0.884 & (0.787,0.993) & 0.04 & 2.146 & (1.885, 2.443) & $<$0.0001\\
Dead & 0.997 & (0.888, 1.120) & 0.96 & 2.205 & (1.937, 2.510) & $<$0.0001\\
H$_0$(t)$^d$ & 1.416 & (1.235, 1.624) & $<$0.0001 & 0.641 & (0.549, 0.747) & $<$0.0001\\
\hline
\multicolumn{7}{l}{$^a$Odds ratio.}\\
\multicolumn{7}{l}{$^b$95\% confidence interval.}\\
\multicolumn{7}{l}{$^c$p-value.}\\
\multicolumn{7}{l}{$^d$Baseline cumulative hazard.}\\
\end{tabular}
\end{center}
\end{table}

\begin{table}
\small
\caption{Data Analysis: Results of Cox regression estimation}
\begin{center}
\begin{tabular}{lccccccccc}
\hline
     & \multicolumn{3}{c}{CC}& \multicolumn{3}{c}{AIPW} & \multicolumn{3}{c}{NNMI(5,0.8,0.2)} \\
\hline
Variable & HR$^a$ & 95\% CI$^b$ & p$^c$  & HR & 95\% CI & p & HR & 95\% CI & p\\
\hline
Age              & 1.015 & (1.012,1.017) & $<$0.01  & 1.015 & (1.012,1.017) & $<$0.01 & 1.018 & (1.015,1.020) & $<$0.01 \\
Black            & 1.437 & (1.286,1.605) & $<$0.01  & 1.436 & (1.276,1.615) & $<$0.01 & 1.442 & (1.307,1.591) & $<$0.01 \\
Others          & 0.887 & (0.781,1.007) &       0.06  & 0.886 & (0.776,1.011) &       0.07 & 0.879 & (0.785,0.983) & 0.02           \\
NoRadiation & 1.056 & (0.978,1.140)  &       0.16  & 1.056 & (0.981,1.137) &       0.15 & 1.044 & (0.976,1.118) & 0.21          \\
NoSurgery   & 1.893 & (1.755,2.042)  & $<$0.01  & 1.894 & (1.756,2.042)  & $<$0.01 & 1.896 & (1.773,2.028)  & $<$0.01 \\
HER2           & 0.940 & (0.867,1.020) &       0.14  & 0.958 & (0.874,1.049)  &       0.35 & 0.939 & (0.864,1.021) & 0.14          \\
\hline
\multicolumn{10}{l}{$^a$Hazard ratio.}\\
\multicolumn{10}{l}{$^b$95\% confidence interval.}\\
\multicolumn{10}{l}{$^c$p-value.}\\
\end{tabular}
\end{center}
\end{table}

\newpage
\begin{table}
\caption{Monte Carlo Simulation Study: Estimation of Cox regression with dependent censoring, where $T \sim  Expoenetial [e^{ln(2)X-ln(2)Z}]$, $C \sim Expoenetial [e^{-2X+0.1Z}]$, $X \sim Bernoulli (0.5)$ and $\delta_x \sim Bernoulli[p(Z,\delta_t,Y)=\frac{1}{1+e^{1.5+0.5Z-2Y}}$. Censoring rate: 0.35; Missing rate: 0.63.}
\begin{center}
\begin{tabular}{lccccccccc}
\hline
     & \multicolumn{4}{c}{$\beta_x=ln(2)$} & \multicolumn{4}{c}{$\beta_z=-ln(2)$} & \\
\hline
Method  & Est$^a$ & SD$^b$ & SE$^c$  & CR$^d$ & Est & SD & SE & CR & Div$^e$ \\
\hline
        & \multicolumn{9}{c}{$N=200$}\\
\hline
FO &          0.689 & 0.188 & 0.194 & 95.8 & -0.704 & 0.335 & 0.314 & 94.6 &\\
CC &          0.668 &0.339 &0.337 &94.8 &-0.988 &0.587 & 0.541 &90.8 &\\
AIPW$_{11}$ & 0.725 & 0.346 & 0.370 & 95.6 & -0.710 & 0.349 & 0.352 & 95.4 & 0\\
AIPW$_{12}$ & 0.758 & 0.339 & 0.393 & 98.0 & -0.791 & 0.428 & 0.508 & 98.8 & 83\\
AIPW$_{21}$ & 0.725 & 0.343 & 0.370 & 95.8 & -0.710 & 0.350 & 0.352 & 95.2 & 0\\
NNMI$_{11}$ & 0.744 & 0.346 & 0.344 & 94.8 & -0.707 & 0.346 & 0.329 & 94.6 &\\
NNMI$_{12}$ & 0.740 & 0.304 & 0.313 & 95.8 & -0.714 & 0.349 & 0.329 & 94.4 &\\
NNMI$_{21}$ & 0.755 & 0.340 & 0.341 & 94.6 & -0.708 & 0.348 & 0.329 & 95.0 &\\
\hline
    & \multicolumn{9}{c}{$N=400$}\\
\hline
FO &          0.689 & 0.134 & 0.136 & 93.8 & -0.677 & 0.228 & 0.220 & 92.8 &\\
CC  &         0.662 & 0.230 & 0.229 & 94.6 & -0.949 & 0.389 & 0.368 & 87.6 &\\
AIPW$_{11}$ & 0.696 & 0.231 & 0.239 & 95.4 & -0.675 & 0.235 & 0.235 & 93.2 & 0\\
AIPW$_{12}$ & 0.731 & 0.230 & 0.264 & 97.4 & -0.740 & 0.303 & 0.331 & 97.6 & 53\\
AIPW$_{21}$ & 0.699 & 0.231 & 0.238 & 95.2 & -0.675 & 0.235 & 0.235 & 93.4 &\\
NNMI$_{11}$ & 0.705 & 0.240 & 0.235 & 93.4 & -0.675 & 0.234 & 0.228 & 92.6 &\\
NNMI$_{12}$ & 0.713 & 0.218 & 0.217 & 95.0 & -0.678 & 0.236 & 0.228 & 93.0 &\\
NNMI$_{21}$ & 0.706 & 0.240 & 0.233 & 93.6 & -0.675 & 0.235 & 0.228 & 92.8 &\\
\hline
\multicolumn{10}{l}{$^a$Average of 500 point estimates.}\\
\multicolumn{10}{l}{$^b$Empirical standard deviation.}\\
\multicolumn{10}{l}{$^c$Average estimated standard error.}\\
\multicolumn{10}{l}{$^d$Coverage rate of 500 95\% confidence intervals.}\\
\multicolumn{10}{l}{$^e$Number of disconvergences for AIPW.}\\
\end{tabular}
\end{center}
\end{table}

\newpage
\begin{table}
\caption{Monte Carlo Simulation Study: Estimation of Cox regression with dependent censoring, where $T \sim  Expoenetial[e^{ln(2)X-ln(2)Z}]$, $C \sim Expoenetial [e^{-2X+0.1Z}]$, $X \sim Bernoulli [p(Z)=\frac{1}{1+e^{0.25-0.5Z}}]$ and $\delta_x \sim Bernoulli[p(Z,\delta_t,Y)=\frac{1}{1+e^{1.5+0.5Z-2Y}}$. Censoring rate: 0.35; Missing rate: 0.63.}
\begin{center}
\begin{tabular}{lccccccccc}
\hline
     & \multicolumn{4}{c}{$\beta_x=ln(2)$} & \multicolumn{4}{c}{$\beta_z=-ln(2)$} & \\
\hline
Method  & Est$^a$ & SD$^b$ & SE$^c$  & CR$^d$ & Est & SD & SE & CR & Div$^e$ \\
\hline
    & \multicolumn{9}{c}{$N=200$}\\
\hline
FO  &         0.693 & 0.188 & 0.195 & 95.4 & -0.685 & 0.334 & 0.317 & 94.2 &\\
CC   &        0.647 & 0.358 & 0.336 & 94.2 & -0.928 & 0.576 & 0.546 & 93.4 &\\
AIPW$_{11}$ & 0.715 & 0.356 & 0.368 & 94.2 & -0.681 & 0.354 & 0.357 & 94.8 & 0\\
AIPW$_{12}$ & 0.737 & 0.337 & 0.395 & 98.6 & -0.756 & 0.452 & 0.529 & 98.2 & 76\\
AIPW$_{21}$ & 0.715 & 0.357 & 0.368 & 95.8 & -0.682 & 0.354 & 0.357 & 94.8 & 0\\
NNMI$_{11}$ & 0.719 & 0.351 & 0.340 & 93.2 & -0.670 & 0.348 & 0.333 & 94.2 &\\
NNMI$_{12}$ & 0.719 & 0.310 & 0.313 & 96.2 & -0.680 & 0.350 & 0.333 & 93.6 &\\
NNMI$_{21}$ & 0.732 & 0.339 & 0.331 & 93.8 & -0.674 & 0.345 & 0.333 & 95.0 &\\
\hline
    & \multicolumn{9}{c}{$N=400$}\\
\hline
FO  &         0.686 & 0.139 & 0.136 & 95.2 & -0.684 & 0.227 & 0.220 & 94.2 &\\
CC  &         0.657 & 0.241 & 0.229 & 93.2 & -0.944 & 0.391 & 0.369 & 87.8 &\\
AIPW$_{11}$ & 0.697 & 0.233 & 0.236 & 95.8 & -0.685 & 0.232 & 0.238 & 94.6 & 0\\
AIPW$_{12}$ & 0.723 & 0.234 & 0.267 & 97.2 & -0.749 & 0.282 & 0.340 & 97.6 & 40\\
AIPW$_{21}$ & 0.697 & 0.234 & 0.236 & 95.4 & -0.685 & 0.231 & 0.238 & 94.6 & 0\\
NNMI$_{11}$ & 0.697 & 0.233 & 0.235 & 95.4 & -0.680 & 0.227 & 0.231 & 95.0 &\\
NNMI$_{12}$ & 0.703 & 0.212 & 0.217 & 95.6 & -0.686 & 0.230 & 0.230 & 95.2 &\\
NNMI$_{21}$ & 0.701 & 0.231 & 0.232 & 96.0 & -0.681 & 0.227 & 0.230 & 94.8 &\\
\hline
\multicolumn{10}{l}{$^a$Average of 500 point estimates.}\\
\multicolumn{10}{l}{$^b$Empirical standard deviation.}\\
\multicolumn{10}{l}{$^c$Average estimated standard error.}\\
\multicolumn{10}{l}{$^d$Coverage rate of 500 95\% confidence intervals.}\\
\multicolumn{10}{l}{$^e$Number of disconvergences for AIPW.}\\
\end{tabular}
\end{center}
\end{table}

\end{document}